\documentclass[12pt,preprint]{aastex}

\slugcomment{Accepted for publication in PASP}

\shorttitle{DAS}
\shortauthors{Platais et al.}

\begin{document}


\title{Deep Astrometric Standards (DAS) and Galactic Structure}


\author{Imants Platais, Rosemary F. G. Wyse}
\affil{Department of Physics and Astronomy, Johns Hopkins University,
    Baltimore, MD 21218}
\email{imants,wyse@pha.jhu.edu}
\and
\author{Norbert Zacharias}
\affil{U.S. Naval Observatory, 3450 Mass. Ave. NW, Washington DC 20392}
\email{nz@usno.navy.mil}




\begin{abstract}
The advent of next-generation imaging telescopes such as LSST and
Pan-STARRS has revitalized the need for deep and precise reference
frames. The proposed weak-lensing observations with these facilities
put the highest demands on image quality over wide angles on the sky.
It is particularly difficult to achieve a sub-arcsecond PSF on stacked
images, where precise astrometry plays a key role. Current astrometric
standards are insufficient to achieve the science goals of these
facilities.  We thus propose the establishing of a few selected deep
($V$=25) astrometric standards (DAS). These will enable a reliable
geometric calibration of solid-state mosaic detectors in the focal
plane of large ground-based telescopes and make a substantial 
contribution to our understanding of stellar populations in
the Milky Way.  In this paper we motivate the need for such standards 
and discuss the strategy of their selection and acquisition and reduction 
techniques. The feasibility of DAS is demonstrated by a pilot study 
around the open
cluster NGC 188, using the KPNO 4m CCD Mosaic camera, and by  Subaru 
Suprime-Cam observations. The goal of
reaching an accuracy of 5-10 mas in positions and obtaining absolute
proper motions good to 2 mas yr$^{-1}$ over a several square-degree
area is challenging, but reachable with the NOAO 4m telescopes and
CCD mosaic imagers or a similar set-up.  Our proposed DAS aims to 
establish four fields near the Galactic plane, at widely separated coordinates. In addition to their  utilitarian
purpose for DAS, the data we will obtain in these  fields will 
enable fundamental Galactic science in their own right. 
The positions, proper motions, and $VI$ photometry of faint 
stars will 
address outstanding questions of Galactic disk formation and
evolution, stellar buildup and mass assembly via merger events.
\end{abstract}





\section{Introduction}

Searches for the answers to fundamental questions of astrophysics and
cosmology have often resulted in technological challenges and
advances. The drive to understand the nature of dark matter
and dark energy has led to proposals for ground-based telescopes with 
large \'{e}tendue (the
product of aperture, A, and field-of-view, $\Omega$), reaching as high as 250
m$^2$ deg$^2$. Such systems enable deep, 
high-cadence and throughput, multi-band imaging of
the visible sky up to $3\pi$ steradians in area  -- crucial
not only to studies in cosmology but also of supernovae, faint 
optical transients, and small bodies in the Solar system.

The success of large digital sky surveys such
as 2MASS and SDSS has shown we can cope with large dataflows and databases, and pointed to necessary future developments, while  recent advances in complex detector designs
\citep{gro00}, have made ever-larger focal plane arrays feasible. 
The most advanced facilities, in terms of their actual implementation,
are the Pan-STARRS \citep{kai04,hod04} array of 1.8m wide-field telescopes
and the  8.4m Large Synoptic Survey Telescope \citep{cla04} (LSST). 
A distinctive feature of each of these observatories is a large field-of-view
(7-10 deg$^{2}$), achieved by applying an innovative optical design 
in combination with a huge 1-3 Gpixel camera, in which the
focal plane is close-packed with several hundred solid-state 
detectors, e.g., CCD chips forming a Focal Plane Array (FPA). 
 Each detector, however, is an autonomous unit
with its own characteristics. In order to emulate a single unified
detector, the parameters of each individual unit, including its
exact location and geometric distortions introduced by the optics, must be 
calibrated. 

These calibrations are crucial at least in two major applications: i)
image resampling and stacking and ii) wide-field astrometry (positions
and proper motions).  Both of them are {\it critical} for weak lens
tomography and the requirement to reach $10\sigma$ limiting magnitude
of $V=28$ by co-adding dithered images \citep{tys02}.  The coherent
galaxy shape distortions caused by cosmological weak lensing do not
exceed 1-2\% changes in the ellipticity (a weighted central second
moment) of galaxy images (cf. \citet{wae00}).  If we assume the
average apparent size of faint galaxies to be $\sim3\arcsec$, then the
lensing signal is only on the order of 30 mas or less. Such a small
signal can be easily confused with systematic errors originating from
imperfect knowledge of the PSF shape, or from inadequate correction for
geometric distortions in dithered images. Thus, translating the
acceptable tolerances in the quality of the PSF, the
LSST Focal Plane Array should be calibrated geometrically to a precision of 
$\sim0.5\, \mu$m or better, corresponding to $\sim10$~mas on
the sky.  Measurements of such high precision are not possible to
perform in the laboratory, and therefore the FPA should be calibrated
and monitored astrometrically on the telescope, using star images
as fiducial points of reference.  If accurate celestial coordinates of
such stars are not known, self-calibration techniques
\citep{and03} can, in principle, provide a provisional reference frame,
albeit with an arbitrary scale and orientation. However, the presence
of geometric distortions and a `broken' (discontinuous) 
FPA would require on the order of a 
hundred optimally dithered, overlapping, and rotated frames, to assure
the success of self-calibration.  Further, high level of
instrumental and atmospheric stability is essential while
continuously obtaining the necessary sequences of such images.

In contrast, all that is needed to calibrate any FPA is a few exposures 
of a dense, deep, and externally accurate astrometric
standard or reference frame. Once derived, the calibration constants
are valid over prolonged periods of time, and  can be monitored by
re-observing regularly the same astrometric standard.  Only such
astrometrically flattened frames can then be shifted and co-added without
a loss of precision.

In this paper we describe the status of existing astrometric reference
frames and show that there is a pressing need to set up a few new deep
astrometric standards specifically for the needs of large imaging
telescopes.  The range of magnitudes of the existing high-accuracy
astrometric standards are too bright for the optimal range of LSST or
other facilities with a similar \'{e}tendue. In just 10~s of
integration time, LSST
reaches $V=24$ and saturates everything brighter than $V\sim17$
\citep{tys02}.  Further difficulties arise from the fact that short
exposures are affected substantially by atmospheric noise, which
diminishes only over a longer 30-60~s integration time. With these longer integrations, 
the saturation level would drop to even fainter
magnitudes.

We argue that deep astrometric standards (DAS), together with elements of
the self-calibrating techniques, will then enable the calibration of
the FPA to the required precision level.  
The DAS fields can be observed and completed on the timescale of a few  
years, significantly prior to the launch (June 2011) and
subsequent catalog release at mission end (yr. 2020) from GAIA. Absolute
proper motions are an important aspect of these astrometric standards,
and these will provide essential constraints on stellar kinematics and
Galactic structure models.  At present, transverse kinematic data over
significant parts of the sky are available down to $V\sim18$ with
proper motion accuracies not better than $\sim6$~mas~yr$^{-1}$
\citep{han04,gir04}.  Two deeper ($B\sim22$) proper motion studies
\citep{chi80,maj92}, with proper motion accuracies of around
1~mas~yr$^{-1}$ or better, probe only a small area in a few lines of
sight at high Galactic latitude.  It is fair to say that we do not
know the kinematics of Galactic populations fainter than $V\sim21$,
especially near the Galactic plane.  For operational reasons, the
selection of potential astrometric standards is limited to low
Galactic latitudes, motivated by the need to have a high 
surface density of stars over the sky. Thus, the astrometric standards 
are expected to make a substantial contribution in constraining Galactic 
structure models via very deep starcounts, multicolor photometry and 
proper motions, with an emphasis on the thick and thin disks.

A favorable combination of pressing calibration needs for large imaging
facilities,  unsolved issues of Galactic structure, and availability
of the appropriate instruments makes the idea of deep astrometric standards
both feasible and timely.  We discuss here in some detail all the required steps in establishing
such standards.

\section{Celestial Reference Frames}

\subsection{Primary reference frames}

At a fundamental level,  the International Celestial Reference System
(ICRS) defines the coordinate axes, and is established by the VLBI
positions of 212 defining compact extragalactic radio sources. 
A set of these and additional 505 sources currently constitutes 
the International Celestial Reference
Frame (ICRF) -- the most accurate (source positional errors at a 0.25~mas
level and the frame orientation good to 20~$\mu$as), albeit 
very sparse, reference frame \citep{ma98,fey04}.
At  optical wavelengths the ICRS is represented by the Hipparcos 
Celestial Reference Frame (HCRF) determined by the Hipparcos Catalogue
\citep{esa97}, excluding all stars flagged as known, and suspected
binary and multiple systems.  The accuracy of the 
HCRF is steadily deteriorating, mainly due to accumulating positional 
errors from the errors in proper motions of the stars, and, by 2006, a typical Hipparcos star 
will have a $\sim15$~mas error in its position. Furthermore, the low
density of Hipparcos stars (on average three stars per deg$^{2}$) and
their high brightness ($V\lesssim10$) prevents direct use of
Hipparcos stars as a reference frame in most applications.

The Tycho-2 catalog \citep{hog00}, containing positions and proper motions
for the 2.5 million brightest stars in the sky, provides 
a denser reference frame. The average density of
Tycho-2 stars per deg$^{2}$ ranges from 20 to 150, depending upon the
Galactic latitude. By 2006, the typical positional errors of Tycho-2
stars will be at the level of  15 to 100~mas, with the larger errors
corresponding to the Tycho-2 limiting magnitude at $V\sim12$.

The second US Naval Observatory CCD Astrograph Catalog (UCAC) provides
a dense and accurate reference frame based on Tycho-2 and, hence,
indirectly on HCRF \citep{zac04}. Down to its limiting magnitude at
$r_{\rm UCAC}\sim16$ this catalog yields a factor of $\sim30$ more
stars per deg$^{2}$ than does Tycho-2, and surpasses the precision of
Tycho-2 at $V\sim10$ and fainter. The precision of UCAC positions
is 15-70~mas, depending on magnitude.  Owing to a number of
innovations and the astrometry-driven approch to all stages of the
observations and reductions \citep{zac00,zac04}, the UCAC currently is
the premier source of reference stars at optical wavelengths.

\subsection{Future reference frames}

For large telescopes, direct use of any of the reference frames listed above 
would be problematic, due to the required shortness of the exposures to  
avoid saturation, and due to the relatively low surface density of 
reference stars, thus limiting their number accessible in any one of 
the CCD detectors in the FPA. There are four large programs which are designed
to improve the reference frame, from substantially to dramatically, 
and extend it down to $V\sim20$.

\subsubsection{URAT}

The USNO Robotic Astrometric Telescope (URAT) is the next step in the 
quest for an ever-fainter, denser, and more precise reference frame 
\citep{zac04a}.  This  is a
0.9m astrometric telescope of innovative design and is intended to 
provide  all-sky reference stars at a 5-10~mas accuracy level, and
with about 30~mas errors at the limiting magnitude $r_{\rm URAT}\sim20$.
The proposed link of the URAT coordinates directly to ICRF, in 
combination with block-adjustment reduction techniques, 
\citep{zac92} not only will improve the alignment of the optical frame
but will also provide a much needed zeropoint for absolute trigonometric
parallaxes.  The expected time frame of the URAT catalog release is 2010.
 
\subsubsection{Space missions: GAIA, OBSS, and SIM}

It is commonly acknowledged that the Hipparcos mission
revolutionized astrometry.  At the time, however, the existing
technology limited the depth of the survey and the number of available
targets to about 120,000, and necessitated an input catalog, as
opposed to a true survey. The next generation ESA cornerstone mission
GAIA \citep{per02} takes advantage of many techological
advancements and is designed to obtain very precise positions and other
astrometric parameters down to $V\sim20$ for one billion objects all
over the sky. It should provide a reference frame at a
micro-arc-second ($\mu$as) level, that is by about two orders of magnitude 
better than is currently possible from the ground at optical wavelengths.
The first early data release from GAIA  is not expected before
2014, with the final catalog release at mission-end, around 2020.

The Origins Billion Star Survey -- OBSS \citep{joh04} is one of the
proposals selected by NASA for study within its Astronomical Search for
Origins program.  The OBSS mission has similar goals to GAIA, but 
considers a different observing strategy and has flexibility in
the selected fields to observe down to $V=25$. If approved for flight,
this mission is likely to be scheduled for the post-GAIA timeframe. 

The NASA  Space Interferometry Mission (now the rescoped SIM PlanetQuest) 
has multiple 
scientific goals, ranging from searches for extrasolar planets to probing 
the astrophysics of QSOs.  It will also provide a global astrometric grid at 
a few $\mu$as accuracy level and  similar precision wide-angle 
astrometry for $\sim20,000$ pre-selected stars \citep{mar03}.

\subsection{Secondary reference frames}

The 20th century's monumental efforts to image the sky has lead to
numerous deep photographic surveys in various bandpasses, now
digitized and reduced into catalogs of objects.  The most advanced, in
astrometric terms, in this class is the three-color, two-epoch catalog
USNO-B1.0 \citep{mon03} which provides reference stars down to
$V\sim21$ with 200~mas accuracy at J2000. As noted by
\citet{mon03}, while this catalog represents  a milestone in the processing of
the object detection archive, more (planned) calibrations and verifications
are required for full exploitation of its astrometric potential.

The 2MASS near-infrared All-Sky Catalog of Point Sources 
\citep{cut03}, complete down to $J=15.8$ and yielding 70~mas 
positional accuracy at the epochs 1997-2001 -- the span of 2MASS
observations -- is a very useful source of positions. Similarly, the Sloan Digitized Sky Survey (SDSS)
provides wide-area, one-epoch zonal catalogs accurate to 45~mas down to
$r\sim20$, and to 100~mas at $r\sim22$ \citep{pie03}.  
Such catalogs, though,  cannot be considered as true secondary 
reference frames since the lack of proper motions makes it
impossible to extrapolate the positions to an arbitrary epoch.
Therefore, the US Naval Observatory has launched an initiative
to combine various sources of astrometric and 
photometric data into the Naval Observatory Merged Astrometric 
Dataset (NOMAD, \citet{zac04b}). One of the goals of this program is to 
minimize local and global systematic errors.

\subsection{Astrometric potential of deep surveys}

As shown above, the limiting magnitude of reference stars with good 
existing astrometry is only about  $V=15$, with possible planned extensions to $V \sim 20$. Only larger
(4m) aperture telescopes and longer exposure times ($t_{exp}\gtrsim10$ min)
will allow us to probe several magnitudes deeper.  There are  at least 
four deep optical surveys which reach $V\sim25$ and cover
several deg$^{2}$ on the sky: the NOAO Deep Wide-Field Survey \citep{jan04},
the CFHT12K-VIRMOS survey \citep{fev04}, the CFHT Legacy Survey
\citep{vei01}, and the Deep Lens Survey \citep{wit02}. Essentially 
all of these surveys are tailored to study various aspects of 
external galaxies, large scale structures, and cosmology. As a 
result, the selected fields are located at high Galactic latitudes
in order to minimize the ``contamination'' by intervening Milky Way 
stars and Galactic interstellar extinction. This kind of minimization,
however, is detrimental to astrometry, since astrometric precision is roughly 
proportional to $1/\sqrt{n}$, where $n$ is the number of reference
stars.  Star-count models (e.g.~\citet{gil81,rob03}) predict that the 
cumulative number of stars at high Galactic latitudes 
($b\gtrsim45\degr$) and with $V<20$ is only about 1,000-2,000 stars 
per deg$^{2}$.  

One survey which does get close to the Galactic plane is the CFHTLS
Very Wide Shallow survey of the ecliptic \citep{vei01}. Having
multiple epoch observations in the Sloan $g',r',i'$ filters, it has a
potential to provide accurate positions down to about $V=23$ at low
Galactic latitudes, lines-of-sight that are abundant with stars. Since
this survey is fine-tuned for  Kuiper Belt Object (KBO) searches
and follow-up, the realization of its full astrometric potential is
uncertain.

\section{Deep Astrometric Standards (DAS)}

The basic astrometric parameters of any celestial object are the
position and motion in the adopted reference system and a parallactic
displacement. 
Astrometry
in essence is a relentless process of establishing better and denser
reference frames so that these basic parameters can be determined with
ever increasing accuracy. As shown in the previous section, the existing
or the near future reference frames do not  extend significantly 
beyond $V\sim20$, 
and in certain applications such as imaging with LSST, that is 
a serious limitation.

\subsection{A rationale for DAS: astrometry and science case}

To provide special astrometric support for large optical imaging 
facilities, we propose the Deep Astrometric Standard (DAS) initiative.
In accordance with the projected specifications of these facilities
and the technical capabilities of ground-based, 4m class imaging
telescopes, the DAS are designed to provide high-precision positions 
and photometry over selected circular 10-deg$^2$ areas down to $V=25$.
These standards will serve as primary reference coordinate sets 
to calibrate the Focal Plane Assembly unit for any large-aperture
telescope and to enable positional stability monitoring of the 
individual units of the FPA -- solid-state detectors such as CCD and CMOS. 
The accuracy of DAS is limited by the accuracy of the primary
reference frames at optical wavelengths, i.e.~at the level of 5-10~mas,
 but this  will be improved each time a better global
reference frame becomes  available, e.g., URAT, GAIA.

In accordance with the mandatory operational requirements described 
in the following section, 
the DAS fields are selected towards the Galactic anticenter 
and inner disk. The photometry and proper motions as deep as $V=25$
in these directions should provide new insights into Galactic structure and galaxy evolution.
Towards the anticenter it will trace the edge of the disk, probe 
the disk flare and low-latitude stellar streams. In the direction of 
Sagittarius and Ophiuchus, the thick disk scale-length, the extent of 
bulge, and contribution by the Sagittarius dSph galaxy at faint magnitudes are 
a few key questions  not yet fully answered. 
An unprecedented combination of depth and spatial coverage of the DAS 
fields at low Galactic latitudes is critical  in attempting to decipher
the complex structure of the Galaxy.

\subsection{Strategy in setting up a DAS field}

The current designs of large survey telescopes, e.g. \citep{hod04},
favor a semi-round FOV, with the diameter ranging between $3\degr$ to
$3\fdg5$.  Thus, the upper limit of a proposed FOV is about 10
deg$^{2}$ and that is also the area we propose here for a DAS
field. With the NOAO CCD mosaic imagers at 4m telescopes (FOV$ =
36\arcmin\times36\arcmin$) it requires 37 partially overlapping
pointings to fill in a round 10 deg$^{2}$ FOV (Fig.~1).

There are several operational requirements as to how to set up such
standards optimally, considering that the majority of the world's premier
large imaging facilities (existing and planned) are located within 
geographic latitudes of $-35\degr<\phi<+35\degr$.

\begin{enumerate}
\item
To ensure access from both hemispheres, the fields should be located near 
the celestial equator.
\item
To ensure year-long access, there must be at least two directions
containing the standard fields separated in right ascension by $\sim12$ hrs.
\item 
To minimize the effect of atmospheric refraction,
one astrometric standard must cross the meridian at a zenith distance
$z\leq20\degr$ while another standard at a higher $z$ serves as a back-up
and a check on refraction corrections. Therefore, the optimal configuration 
is a pair of fields at similar right ascensions and located symmetrically 
to within $\pm20\arcdeg$ on both sides from the equator.
\item
A field must be dominated by stars, not galaxies which cannot be centered
as precisely as can stars. Hence, the fields should be near the Galactic 
equator as well.  Note that the celestial and Galactic equators cross 
each other at RA=$6\fh86$ and $18\fh86$ (J2000). 
\item
A field must be free of dark clouds or emission nebulae and should not contain
bright stars ($V\lesssim7$). A fairly uniform and not-too-dense distribution 
of stars across the field is also required.  The POSS photographic atlas 
is convenient in searches for promising directions on the sky. Similarly, 
the Hipparcos Millennium star atlas is useful in locating areas devoid 
of bright stars.
\end{enumerate}

\subsection{First DAS fields}

Guided by the criteria outlined in Sect~3.2 and by the desire to maximize the
science return, we made a selection of four DAS fields (Table 1).
There are two Northern `winter' fields (GTO, Hya) and two
`summer' fields -- Oph and Sgr. Thus, for each hemisphere
there are two primary and two secondary astrometric standard fields.
A visibility analysis for the KPNO and CTIO locations indicates
two one-month windows when no primary standard field is culminating
at night, but that alone does not warrant setting up additional
fields far away from the equator. Two fields cross the ecliptic
plane, where searches for Kuiper Belt Objects are feasible.
One of the toughest requirements is avoiding bright stars in
a 10~deg$^2$ field. In the selected DAS fields the brightest star 
is at $V=6.4$. It should be noted that our minimalistic approach to
the selection of fields is dictated by telescope time limitations.
More deep fields in other directions would provide more constraints to
Galactic models and would also ease the astrometric calibrations.

From the standpoint of interstellar extinction three of our DAS fields are
almost transparent. The mean reddening $E(\bv)$ values are obtained from 
\citet{sch98} maps.  The GOT (Gemini-Orion-Taurus)
field has a much higher mean reddening, 
namely, $E(\bv)\sim 1.4$ which amounts to $\sim4.5$ mag of a total 
absorption in the $V$ bandpass. This field (see \citet{mer98}) contains 
three sparse open clusters having the following reddening and
distance: NGC 2129 ($E(\bv)$=0.6; d=1.5 kpc), Berk 21 ($E(\bv)$=0.7; 
d=5.0 kpc) and Basel 11b ($E(\bv)$=0.3; d=1.7 kpc), thus indicating 
that the light absorbing material is not distributed evenly and
there very well could be windows with lower absorption.

Star-count models are most reliable away from the Galactic plane,
where extinction is lower and the stellar distribution is smoother.
The Besan\c con model adopts a different approach, one of stellar
population synthesis \citep{rob03}.
We obtained the expected number of stars per deg$^{2}$ in our proposed fields using both a straightforward star-count model \citep{gil81} 
and the Besan\c con population synthesis model.  These are  
indicated by the last two columns in Table~1, where good agreement is seen apart from close to the disk plane. Guided by these apparent
stellar densities it may appear that crowding could be severe. 
Therefore in November 2004 we obtained a series of $VI$ exposures in the GOT 
field ranging from 10~s to 900~s with the NOAO 4m telescope and
CCD mosaic imager (FOV$=$0.36 deg$^{2}$). From the long exposures, 
the average number 
of unsaturated and well-centered images (centroid $\sigma\sim0.05$ pix) 
over the entire CCD mosaic is 16,000 or well below the acceptable level
of crowding.
Extrapolating smoothly over the entire 10~deg$^{2}$ leads to an expected 
total number of stars down to V=25 in the GOT field of 
$\sim500,000$.  This is significantly lower than the model predictions.  A trial 10~min exposure of the Sgr field in the $V$ bandpass
indicated $\sim60,000$ stellar images over the entire CCD mosaic and no
substantial crowding across the field.  
This high a number of stars with precise positions should
provide an excellent reference frame to support astrometric calibration 
of any CCD mosaic in existence or in the planning stages. Despite  
the preliminary status of the measured stellar densities, we note
large discrepancies between the model predictions themselves and between 
either model and the actual starcounts. The DAS program is well-suited
to resolve these differences towards the selected directions, although
it is clearly not optimal for the entire Galaxy due to the small number 
of fields.

The Galactic coordinates of the selected fields show that the Oph and Sgr
fields probe the Galactic inner disk, while the GOT field is close to 
the Galactic
anticenter, and the Hya field is closer to the direction of Galactic
rotation. It is instructive to review what these directions can offer
for studies of Galactic structure and kinematics.

\subsection{Galactic structure studies with DAS fields}

These astrometric standard fields are expected to make a substantial
contribution to constraining Galactic structure models and our
understanding of the Milky Way stellar populations.  Each of the four
fields contributes to a specific aspect, and taken all together will
provide constraints on population gradients and substructures. Our
individual fields are large enough to probe gradients on $\sim$~kpc
scale -- at a fiducial distance of 10~kpc ($V=20$ and $M_V=+5$), 
three degrees on the sky corresponds to $\sim 500$~pc -- and more
distant stars probe even larger spatial extent.
The deep photometric data will allow
star-count analyses, e.g., \citet{sie02} to constrain the particularly
poorly known radial distributions of the thick and thin disks. The
depth of these fields at $V=25$ would allow the detection of G-dwarfs
($M_V=+5$) out to 100~kpc, with no extinction, and out to 16~kpc in
the more heavily reddened GOT field.  Even more importantly, the
accuracy of our DAS absolute proper motions is expected to be $\sim
2$~mas/yr, a factor of three better than the quoted rms error of the
photographic-based Lick proper motion catalog (Hanson et al.~2004),
and only a factor of two below the 4-m (photographic) survey of
\citet{maj92} which used a 16~year baseline and which was limited to
$B \sim 22.5$, some 1.5~magnitudes brighter than our limiting
magnitude.  Our data will provide the deepest available absolute 
proper motion data over relatively large areas of the sky. They
will enable diverse science projects such as refining the kinematical
properties of Galactic stellar populations \citep{maj92,men00},
identification of debris from Galactic mergers \citep{maj96}, and
constraints on the surface mass density of the disk.  The Reduced
Proper Motion Diagram (RPMD) would provide nearly distance-independent
classification of Galactic populations that incorporates kinematics.
Stars can be classified by their location on this diagram
(e.g. \citet{sal02}), greatly facilitating the identification and
characterization of Galactic stellar components, such as the thick
disk, first kinematically identified through this technique
\citep{wys86}, and the Galactic halo \citep{gou03}.  
The fields will be particularly useful for
investigations of the Galactic disk, which has gained new interest as
both recent observations, indicating unexpected substructure
(e.g., \citet{new02}) and recent theory, indicating significant
mergers into the thin disk (e.g., \citet{aba03}), have emphazised
our lack of knowledge of the disk far from the solar circle and its
importance in constraining theories of disk galaxy formation and
evolution.

At faint magnitudes, accurate star-galaxy separation is crucial to
the interpretation of starcounts. As shown by \citet{rei96}, at high galactic
latitudes ($b>45\degr$) the surface densities of stars and galaxies are equal
at $I\sim18.5$, and at $I\sim24$ galaxies outnumber stars by a factor of
$\sim40$. The DAS fields are located at low and moderate galactic
latitudes, therefore the contamination ratio of galaxies is reduced by 
a factor 6-75 compared to high Galactic latitudes.  
Additionally, background galaxies
tend to be bluer than low-mass disk stars, which are the dominant
contributor in our four fields.  A combination of object morphology 
and CMD analysis is expected to provide a reliable discrimination 
between stars and galaxies, with proper motions greatly helping.

\subsubsection{The outer edge of the Galactic disk}
The GOT field is conveniently positioned to probe the outermost regions 
of the disk in
the anti-center direction. At 8~kpc, a $3\fdg5$-diameter field 
probes a $\sim 500$~pc wide spatial cone and we should therefore be
able to constrain possible variations in thin-disk scale-height
(i.e.~flaring) in addition to determining the disk scale-length
(cf. \citet{rob92a}) and investigating the putative disk `edge' at
$\sim6$ kpc from the Sun \citep{rob92b}. These latter
results were based on star counts to a similar depth as we propose
($V=25$), but over a much smaller area, only $0.008$~deg$^2$. 
The direction of the Galactic anticenter towards Taurus is
strewn with dark nebulae and star forming regions that have a high and
variable extinction. The much larger  area (a factor of $\sim1000$) of 
our survey  provides an
opportunity to map out extinction and find new transparency windows
through which to measure the disk parameters.

\subsubsection{The Stellar Warp in the Disk} 

Another aspect of our ignorance
about the outer Galactic disk is the amplitude and shape of the warp
in the stellar disk of the Milky way.  Does it follow that in the gas?
Do old stars and young stars exhibit different warp structure? 
The warp has taken on new importance recently due to the controversy about the
role it could play in the overall non-axisymmetric structure in the
stellar disk.  There have been claims and
counter claims as to whether or not the stellar warp has parameters
sufficient to mimic a distinct overdensity such as has been identifed
as the core of the proposed Canis Major dwarf galaxy at $\ell = 240\degr,
\, b = -7\degr$ \citep{bel04,mar04a,mar04b,mom04}.
Our Hya field probes the
corresponding northern latitudes to these fields, where the warp
should manifest itself as a lack of distant disk stars. Again, our fields
are sufficiently large that we will be able to constrain the warp
parameters by star counts across the field.

\subsubsection{The surface mass density of the disk} 

Our outer disk fields can
be used to facilitate the estimation of the surface mass density of
the disk beyond the solar neighborhood.  Disk dwarfs can be identified
through the reduced proper motion diagram, and spectroscopic follow-up
will allow the determination of metallicities and thus photometric parallaxes, together
with full 3-D space motions.  The vertical motions may then be
combined with vertical star counts in an analysis of the total surface mass density  in
the disk (cf. \citet{kui89}).

\subsubsection{Substructures in disk \& halo}  

Our Hya and GOT fields will also
shed more light on the `ring' that apparently encompasses the Galaxy
\citep{iba03}, seen most prominently in the anticenter direction
\citep{new02,yan03}.  Is this a feature in the
disk, or a remnant of a shredded galaxy, such as the Canis Major dwarf
discussed above?  Our large fields and widely separated lines of sight
to the outer disk will constrain both small-scale and larger-scale
variations, particularly when combined with imaging data from the
Sloan Digital Sky Survey and its extension SEGUE \citep{bee04}.

Our Sgr field is located at a prominent tail from the Sagittarius
dwarf spheroidal, seen clearly in M-giants (see Fig.~3 in \citet{maj03}).
A detailed knowledge of stellar population in the tail
will better constrain models of the dynamical interaction. 

Our expected proper motion error of 2~mas/yr 
at a distance of one kpc translates into a transverse velocity error of 
10~km/s, and at  distances larger than about 10~kpc, the transverse 
velocity error is large enough to make even statistical separation of 
populations  by proper motions alone difficult. This problem is 
substantially mitigated by applying the Reduced Proper 
Motion Diagram, which will be used in the analysis.
Indeed, our deep photometry and precise proper motions allow the
derivation of reduced proper motion diagrams with low enough errors
that a distance-independent identification of substructure in this
plane may be carried out.  We note that at our faint magnitudes, $V
\gtrsim 19$, giants with absolute magnitudes $M_V \sim -1$ are at
distances of greater than 100~kpc, the very outer limits of the
stellar halo of the Milky Way (e.g.~as traced by RR Lyrae in the Sloan
Digital Sky Survey, \citet{ive04}). Our analysis will therefore
be using main sequence stars, the dominant stellar populations, and
probe distances from several kiloparsecs ($V$ $\sim 19$) to the edges 
of the stellar halo ($V$ $\sim 25$).  One should note that
kinematic signatures of substructure are not limited to the outer halo
where dynamical times are longest -- on the contrary, there is a
wealth of local (less than 1~kpc) `moving groups' (e.g. \citet{fam05,hel05}) 
and the challenge is to identify their
origins -- some clearly are better interpreted as due to dynamical
perturbations in the local disk. Extending our knowledge with fainter
stars, such as could be achieved with the dataset proposed here, would
clearly be beneficial, albeit that our pencil-beam approach, as
opposed to all-sky, provides more limited constraints. \citet{din02}
demonstrate the power of pencil-beam proper-motion surveys, with data that 
are shallow by the standards of the survey proposed here, but more precise.

\subsubsection{Halo/bulge interface} 

The Oph field, and the Sgr field, probe
the halo/bulge interface and the inner disk along the
line-of-sight. Much recent interest concerns the nature of
`pseudo-bulges' which have exponential surface brightness profiles,
rather than the canonical $R^{1/4}$ de Vaucouleurs' profile (e.g.
\citet{wys99,kor04}). A possibility is that they result
from long-timescale instabilities in the inner disk, associated with
the formation and destruction of bars.  The Milky Way bulge is
apparently such a `pseudo-bulge' and the data from our deep fields
here will allow investigation of the similarity or otherwise of the
stellar populations in the inner disk and bulge, necessary for such
secular evolution models of bulge formation.  The further decomposition of  the
central regions into `halo' and `bulge' is important in understanding
formation scenarios for the Galaxy.

\subsubsection{Low mass \& low luminosity objects}  

Our wide, deep fields
provide an opportunity to study rare, faint objects such as white
dwarfs and L and T brown dwarfs.  Due to the intrinsic faintness of L
and T dwarfs it is essential to have a combination of depth, color and
{\it kinematic\/} information just to identify these low-mass Galactic
constituents.  The same holds true for the more massive white
dwarfs. The luminosity functions of each of these objects is still uncertain,
at best.  Scaled to the CFHT Legacy Survey estimates
\citep{vei01}, the DAS will
be able to probe the population of L dwarfs out to $\sim250$ pc. Much
dimmer T dwarfs can be traced out to 30 pc.  Our white dwarf
candidates will be identified by their reduced proper motions,
allowing spectroscopic follow-up. 

\subsection{Kuiper Belt objects}

To maximize the science return, we may want to explore
the two areas where all three planes --  equatorial, Galactic, and ecliptic -- 
get close together. Since the searches for Kuiper Belt objects (KBO)
such as Deep Ecliptic Survey \citep{mil02} partially avoid the Galactic 
plane, due the crowding and the presence of bright stars,
DAS should be able to fill in this gap. In addition, DAS will go 
$\sim0.5$ magnitudes deeper than will the Deep Ecliptic survey and will provide
$VI$ photometry for all detected KBOs. The compilation by \citep{hai02}
indicates a wide range of colors for minor bodies in the outer Solar
System ($0.5<V-I<1.8$) while the Sun has $V-I=0.69$. 
However, the statistics are still poor since only
$\sim10\%$ (or about 150) of the known KBOs have their color measured. 
Scaling from the current KBO discovery rate with NOAO 4m telescopes \citep{mil02}, 
we can expect 20-30 new KBOs in each of the GOT and Sgr fields. 

\section{Instrumentation and Reduction Techniques}

With CCD detectors there are three basic modes to obtain an image. Most
common is the guided stare-mode, identical to photographing
the sky. Alternatively, one may either stop the telescope
and let the sky drift across a CCD (drift-scanning mode) or drive
the telescope at a rate that is synchronous with the charge transfer rate
applied to a CCD (TDI -- time delay \& integrate mode) as in the SDSS 
\citep{pie03}. In these last two modes, parts of a sky image are formed 
sequentially 
in time, which leads  to a partial loss of information about the atmospheric
noise -- a key factor limiting the accuracy of ground-based wide-field
astrometry. In the following section we consider only the stare-mode.

The coordinates of astronomical objects in the frame of a detector
always contain instrumental effects mainly caused by the imaging
optics, known as geometric distortions.  In the case of ground-based
observations, on top of these distortions one must deal with the
effects of atmospheric refraction, which changes the path of incoming
light as a function of zenith distance and wavelength.  At a
microscopic level, atmospheric refraction is highly variable and has
very short temporal and spatial coherence intervals.  Ordinary CCD
devices operating in a static light-integration regime cannot account
for such high frequency and spatially unstable effects, other than to
expect that longer integration times tend to average them
out. Indeed, both theoretical predictions \citep{lin80} and
empirical relationship, e.g., \citet{zac96} indicate that the uncertainty 
in positions due to atmospheric noise is proportional to $t^{-\onehalf}$,
where $t$ is the integration time.
However, there is a fine line between the desire to extend the
integration time and the danger of image saturation. As a result, one
usually ends up with a less than optimal exposure time (too short)
which inevitably includes a substantial non-modellable contribution of
atmospheric noise which limits the astrometric accuracy.

The first step in astrometric reductions is to derive distortion-free
coordinates. This normally involves some external reference frame
although there is a way to self-calibrate the distortion, at the
expense of an unknown absolute scale factor \citep{and03} and with
a caveat that the concept has been proven for the space-based
HST observations only.
Here we will consider the case with reference frames, first in a single
CCD chip regime and then with a CCD mosaic.

\subsection{Single CCD detector}

Astrometry with single CCD chips is very similar to astrometry with a
photographic plate.  First, the celestial coordinates of reference
stars are converted into standard coordinates via the gnomonic (TAN)
projection of a sphere onto the plane at a tangent point.  Then, the
measured Cartesian coordinates of the reference stars are adjusted to
their standard coordinates using least-squares as a maximum likelihood
estimator. This step involves a plate model which, besides coordinate
offset, rotation, and scale factor, should also adequately represent 
all other significant effects such as tilt and geometric distortions. 
Finally, the calculated standard coordinates are projected back onto 
the celestial sphere. In certain
applications, and in the case of poor sets of reference stars,
it is desirable  to pre-correct the measured coordinates for tilt and
distortions, if they are constant. 

A couple of potential problems may arise in CCD astrometry. First,
the field-of-view may be too  small and  not contain a sufficient number
of reference stars. 
Second, the images of brighter stars, which typically are the best reference 
stars, may be saturated and unsuitable for precise measurements. In
either case, the plate model parameters are not satisfactorily 
constrained. In other words, the accuracy of the resulting celestial
coordinates could be much lower than the precision of the measured Cartesian
coordinates.  Differential astrometry employing the measured
coordinates can yield very high precision. Thus, \citet{pra96}
have achieved a 1~mas precision over a few arc-min field. Similarly,
the USNO parallax program with the 1.55~m telescope routinely
produces relative astrometric measurements accurate to 3~mas 
per epoch \citep{dah02}, which then lead to a sub-mas precision in
parallaxes over several years of observations. 
These quoted values though must be qualified by the caveat
that extreme care was taken to minimize all potential sources
of systematic and random errors.

This classical astrometric reduction scheme certainly has limitations
stemming from a limited precision and possible systematic errors 
in the reference star catalogs. Any position, magnitude, or color 
dependent error in the reference stars will propagate into the
distortion coefficients and target star positions.

\subsection{CCD mosaic device}

Large CCD mosaic devices for direct imaging are now available for many 
telescopes of different sizes \citep{gro00}, and they indeed dramatically
increase the productivity of these facilities. However, we cannot
ignore the fact that an array of CCDs is a fragmented detector
consisting of separate units, each with its own characteristics.
For astrometry, it is extremely inconvenient to have a `broken-up' focal
plane assembly. Therefore, it is not surprising that, for instance, 
at the NOAO 4m Blanco and Mayall telescopes a common practice 
is to extend the concept of one-chip astrometric reductions to each 
individual chip in the mosaic \citep{dav98}.  
The NOAO Mosaic Imagers include  eight thinned, back illuminated, 
2K$\times$4K SITe CCDs. 
Thus, each solution, or a set of plate constants in the FITS header, contains 
information on the optical distortion and the chip location relative to the
telescope's optical axis. The plate constants are determined at the beginning
of an entire run, during which the telescope zeropoint may change, and
hence a repeat off-line solution with USNO-B or UCAC stars 
safeguards against possible shifts in predicted coordinates, cf.
 \citep{mil02}.  

The obvious simplicity and convenience of chip-oriented solutions 
nonetheless neglect two basic facts: 
1) the bulk of geometric distortions produced by the
optical system is axicentric and can be quantified by a {\it single} set of
constants for the whole CCD camera; 2) there is only one projection 
of the celestial sphere onto the focal plane at a single tangent point.
These two properties essentially demand unification of the fragmented detector
into a single superplate, characterized by a common Cartesian coordinate
system, $x,y$. Such a superplate then allows us to enlarge the number 
of reference stars by a factor equal to the number of CCD chips and 
to {\it reduce} the number of unknown constants by nearly the same factor. 
A set of the so-called chip constants --
the $x_i,y_i$ position of the $i$-th chip center, a rotation angle, $\Theta_i$,
around this center, and a look-up table of higher-frequency distortions
emanating from imperfections of the CCD surface (non-planarity, small tilt,
and twisted/skewed pixel axes) --  then fully describe each chip's metric 
\citep{pla02}. 

Perhaps one of the earliest efforts to develop the superplate concept
is the work by \citet{kai99}, later implemented by the
Terapix\footnote{\url{http://terapix.iap.fr}} 
data reduction center aimed at complete and automated reductions of 
large datasets such as the output from the CFHT 3.6m 
telescope with the MegaPrime CCD mosaic imager.  To better understand 
the astrometric properties of CCD mosaics, \citet{pla02} examined in 
great detail the NOAO CCD Mosaic Imager by using an astrometric standard 
field. The summary of this study is given in the following section.

\subsubsection{The techniques}

The technique of deriving the metric of a CCD mosaic using an
astrometric standard is straightforward -- the measured pixel
coordinates of reference stars, $X_p,Y_p$, must be adjusted to their
tangential coordinates, $\xi,\eta$, calculated via the gnomonic
projection. Following the concept of a superplate, the pixel
coordinates, $X_p,Y_p$, should be translated into the global Cartesian
CCD coordinate system $x,y$. That can be done indirectly through the
reference stars and a least squares adjustment
$x,y\Longrightarrow\xi,\eta$, employing an appropriate polynomial
plate model (including all distortion terms).  Consider that we have
derived a set of approximate values for the chip centers and rotation
angles $cx_i,cy_i,\Theta_i$ and, hence, the global coordinates
$x,y$. Usually, the initial chip centers are estimated from the
measured chip separations but rotation angles can be safely assumed to
be zero. Apparently, the adjustment $x,y\Longrightarrow\xi,\eta$ with
such crude global coordinates will produce very poor residuals,
reflecting our first guess for the chip constants.  In order to
improve them, we should select a fixed set of reference stars and
minimize the standard error, $\sigma$, of the adjustment
$x,y\Longrightarrow\xi,\eta$ while searching through the parameter
space $cx_i,cy_i,\Theta_i$ at a fixed $i$. In other words, an
adjustment, equivalent to $\chi^2$, is used to find iteratively an
$i$-th triad of the chip constants $cx_i,cy_i,\Theta_i$, one at a
time. It is the minimum of the standard error $\sigma$, 
that signals that
the chip constants have been found. Normally, the next step is to find
the distortion center and refine the geometric distortion terms, using
the same adjustment but now adopting the obtained mean chip constants
\citep{pla02}.
 
In practice, the chip constants derived in this way are sensitive to 
the achieved accuracy of geometric distortion determination.
Since geometric distortions are 
stable over the timespan of fixed optical adjustment of a telescope,
it is desirable to pre-correct the pixel coordinates $X_p, Y_p$ for
geometric distortions, provided the optical center (tangent point)
is known or can be found. This allows simplification of the plate model
down to linear terms only. The repeated adjustments of the values of the chip constants
with a linear model provide more accurate chip constants. The accumulated 
residuals from these adjustments then allows one to generate a look-up 
table of the higher-frequency, unmodelled, distortions.  

Limited accuracy and low density of reference stars are some of the 
other factors that may affect the precision of the chip constants. 
If this is a serious problem,
then self-calibration techniques can provide a solution \citep{and03}. 
This approach 
makes use of {\it all} images in the overlapping area. The advantage
of self-calibrating techniques is a higher number of stars and the
presence of only one source of random errors, that is the errors
from image centering. The disadvantage is a large number of 
overlapping frames required to achieve high precision and an inability
to derive the absolute scale.

The strongest test this technique has faced so far is in calculating 
proper motions in the open star cluster NGC 188 from a combination of 
old photographic plate measurements and the CCD mosaic data \citep{pla03},
which resulted in a 0.15 mas yr$^{-1}$ accuracy and indicated no 
apparent systematic errors.

\subsubsection{A pilot study around NGC 188}

As a by-product of the WIYN Open Cluster Study (WOCS), we have created 
a relatively deep ($V\leq21$) astrometric standard in the 0.75 deg$^2$  
area around the open cluster NGC 188 \citep{pla03}. It is based upon
30 old photographic plates from assorted large-aperture telescopes, in
combination with more than a 100 CCD mosaic frames obtained at the 
KPNO 4m telescope with the NOAO CCD Mosaic Imager.  
The plate measurements and the global CCD coordinates 
$x,y$ of each frame, as described in Sect 4.2.1, were mapped into the 
specifically constructed \citep{pla02} Lick intermediate 
catalog of positions and proper motions around NGC 188. 
The Lick intermediate catalog is similar to UCAC in terms of limiting magnitude 
at $V\sim17$, positional precision ($\sim60$~mas or better), and 
proper motion accuracy (2-7~mas~yr$^{-1}$). 

The WOCS astrometric standard for optimally exposed 
isolated stars yields a 2~mas precision in positions.  At the time it 
was believed that the positions are within $\sim$5-10~mas on the system 
of ICRS \citep{pla03}.  However, a detailed comparison with 
the preliminary UCAC data in that area of the sky 
shows much larger systematic
positional differences.  There is a clear spatial trend in the 
differences `UCAC-WOCS' indicating a scale problem (Fig.~2). 
In other words, a $\sim6.5$~mas per arcmin correction
in both axes would bring the two systems of coordinates to a nearly 
perfect match.  The observed systematic differences are 
unexpected since in both cases the astrometric reductions were 
done using the Tycho-2 catalog, albeit utilizing different 
sub-samples and, in the case of WOCS standard, indirectly via the Lick
intermediate catalog.
It should be noted that the Lick intermediate catalog, used as
a reference frame in \citet{pla03}, contains  magnitude-dependent
systematic errors.  This is demonstrated by direct coordinate
differences `Lick-WOCS' (Fig.~3,4). The common Tycho-2 stars, indicated
by the filled squares, show a considerable offset (up to $\sim90$ mas) 
between the two catalogs. Additional tests, involving the UCAC and 2MASS
positions, show that it is the Lick intermediate catalog 
responsible for this kind of systematic error.  Nevertheless, 
such magnitude-dependent systematic errors alone should not introduce 
a spatial trend in the coordinates of the WOCS astrometric standard. 
At this point we cannot identify unambiguously the source of the spatial trend. 

A large number of CCD Mosaic frames around NGC~188 allows us to
obtain some statistics on the residual scatter as a function of FWHM
and exposure time. The coordinates of all frames were subtracted from
the final catalog \citep{pla03} and, after some trimming, the mean
dispersion, $\sigma_{\rm pos}$, was calculated for each exposure.
As indicated by Fig.~5, there is a strong correlation between
the FWHM of images and $\sigma_{\rm pos}$. The dependence on exposure
time is almost nonexistent at FWHM$<6$ pixels.  Apparently,
a 30 s exposure is already sufficient to reach the floor of dispersion 
distribution at $\sim20$ mas, at least for the zenith distances
and seeing conditions prevailing at the time of these observations.

This case underscores the uncertainty in linking a catalog to the ICRS, 
at least over a small area ($\leq1$ deg$^2$). 
It should also be noted that, as a circumpolar object, NGC 188
is always at large zenith distance ($z>53\arcdeg$ from Kitt Peak) and over
60\% of all frames were taken with short exposures ($\leq30$ s),
thus were exposed to deleterious atmospheric effects on astrometry.
With longer exposures, a near zenith pointing, and in sub-arc-sec seeing
high positional precision can be achieved with a substantially smaller 
number of CCD mosaic frames -- as planned for the DAS observations.

\subsubsection{Astrometry with the Subaru Suprime-Cam imager}

In the context of LSST it is instructive to consider a telescope with 
a similar aperture to LSST. The Subaru 8.2m telescope is the only one in its 
class having a mosaic CCD imager with large FOV ($34\arcmin\times27\arcmin$). 
We used the SMOKA Science 
archive\footnote{\url{http://smoka.nao.ac.jp}} and extracted ten 
short exposure Suprime-Cam CCD mosaic \citep{miy02} frames of the 
ESP field 1 in the Sloan $i'$-filter taken in May 7, 2002 \citep{mon04}. 
The corresponding sub-frame identificators span SUPA00106300-SUPA00106399.
On these frames, more than 400 UCAC stars can be identified. Many of them 
are overexposed but surprisingly without a substantial degradation in
the positional precision.
Following the notation in \citet{pla02}, the original pixel coordinates,
$X_p$ and $Y_p$ (D. Monet, private communication) were used in deriving
preliminary chip constants and the optical field angle distortion (OFAD)
parameters. The geometric center for the entire CCD frame was adopted
at $x_c$=5300 and $y_c$=4100 pixels. The derived chip constants are
given in Table~2, which contains the chip number, the chip constants,
$dx$ and $dy$, in pixels and the rotation angle, $\Theta$, in radians.
The corresponding CCD layout is given in Fig.~6.
The standard error estimates, $\epsilon_{dx}$, $\epsilon_{dy}$, and
$\epsilon_{\Theta}$, are for a single determination of the chip
constants. It should be emphasized that these constants are valid
only for the epoch 2002.3, and extrapolating to other epochs 
requires additional studies on the stability of the chip constants. 

The cubic distortion term in the $i'$ bandpass is
$(-4.767\pm0.012) \times 10^{-16}$ rad pixel$^{-3}$ in right ascension and
$(-4.674\pm0.017) \times 10^{-16}$ rad pixel$^{-3}$ in declination.
The fifth order term along the same axes is 
$(+2.11\pm0.03) \times 10^{-24}$ rad pixel$^{-5}$ and
$(+1.89\pm0.03) \times 10^{-24}$ rad pixel$^{-5}$. 
The reason for providing separate distortion solutions in each
axis is to show their consistency, which may vary from one telescope
to another indicating subtle deviations from a perfect reference
frame and/or optical system itself, and possible pixel scale differences
along the $x$ and $y$ axes. For practical applications,
distortion solutions should be averaged.

We note that the 5th order term indicates the presence of a barrel-type
distortion while the cubic distortion is the pincushion type.
A similar conclusion can be drawn from the optical distortion
parameters derived in \citet{miy02}, although they used a different
field-distortion model.  The global (whole-frame)
solutions using over 400 UCAC reference stars routinely yielded
a standard deviation of 40~mas or better. The contribution of 
uncertainties in the reference frame is estimated to be around 30~mas,
leaving $\sim20$~mas attributable mainly to unmodelled 
systematic errors in the positions.  The 
Suprime-Cam camera appears to have excellent astrometric properties. 

The main purpose of these astrometric solutions was to explore
atmospheric effects in 10~s and 30~s exposures. All ten sets of 
equatorial coordinates were combined to derive a catalog in
the direction of ESP field 1 (RA=18$^{\rm h}$26$^{\rm m}$,
Dec=$+21\arcdeg42\farcm3$, J2000). The coordinate differences between 
the catalog and a selected CCD mosaic frame should be representative
of atmospheric noise. Apparently, these differences (or residuals)
will also include the modelling error, mainly originating from the 
least-squares adjustment to obtain the equatorial coordinates. 
The contribution of modelling error, however, is kept constant by using 
frames with identical telescope pointing, the same plate model, and 
essentially the same reference stars in all frames. 

A distinct advantage of this approach is the possibility to
probe atmospheric noise over the scales substantially larger than the
chip size. That is clearly demonstrated by Figs.~7,8 in which the 
vectorial pattern is consistent between adjacent chips. 
On average the residual vectors 
from a 10~s exposure (Fig.~7) are almost twice as long as 
from a 30~s exposure (Fig.~8).  It is conspicuous that
these vectors appear to have a prevalent North-South direction, albeit
atmospheric noise should be acting randomly. This unusal phenomenon 
clearly requires further studies. It should be noted that the Suprime-Cam
frames considered in this study have not been obtained under optimal
conditions. For instance, the guiding was off and that resulted in
somewhat asymmetric images. Nevertheless, on the 30~s exposure frames it is
possible to isolate narrow but long, up to $25\arcmin$, stretches of 
the sky showing residual scatter at the level of only 3~mas. 
On the 10~s frames, the extension of such low-residual areas is much
shorter -- on the order of $5\arcmin$ only. This appears to be 
indicative of the area  on the sky where  high-precision
differential astrometry, reaching the intrinsic error floor, is 
feasible with short exposures.
 
\subsection{Limitations in CCD mosaic astrometry}

The major source of uncertainties in this technique is the limited 
positional accuracy of a reference catalog and/or insufficient number 
of reference stars.
The typical FOV of existing large telescopes equipped with a CCD mosaic
is about 0.4 deg$^{2}$. The CFHT MegaPrime camera covers nearly
a full square degree which probably is an upper limit for this 
type of telescope unless a new type of wide-field field corrector can be
manufactured \citep{epp03,kom04}.
Another critical number is the typical centering precision of 0.02-0.05 pixels 
for optimally exposed stellar images on the CCD mosaic \citep{pla02}. That
translates into a 5-15~mas precision for the average pixel size of 
$\sim0\farcs2-0\farcs3$.  How many reference stars of comparable
positional accuracy can we find in a typical square degree on the sky? 
Near the Galactic plane the density of UCAC stars, $n_{\rm U}$,
is indicated in Table~1. In the case of the NOAO 4m telescopes and their 
CCD mosaics that amounts to $\sim$500-1300 stars over the FOV.
If only that many stars are used to derive geometric distortions and
chip locations, all at once, the chances are that the solution may not
provide the desired accuracy. Therefore, additional constraints 
available from the areas of overlapping frames are vital to reach  
a 5-10~mas precision across a larger FOV. We stress here that
the most important aspect of wide-angle astrometry is the
astrometric flat-fielding, so that the transformation of
any two overlapping fields or frames is purely conformal, i.e.,
limited to offset, rotation, and scale only.  The mapping accuracy 
into the ICRS is then entirely dependent upon the degree to which 
a concrete reference frame represents the ICRS.

There are two significant factors that potentially can limit the 
astrometric accuracy with CCD mosaics:

\begin{enumerate}

\item
Atmospheric turbulence puts a fundamental limit on the temporal and 
spatial precision of astrometry in accordance with the predictions from 
\citet{lin80} and the positional variance analysis from  CCD 
observations \citep{zac96}. Thus, over the angular extension
of the CCD Mosaic's chip at the KPNO 4m telescope ($18\arcmin$) 
the standard deviation of the atmospheric noise contribution in a 10~s
exposure is $\sim30$~mas \citep{pla02}.  
The effect of atmospheric noise is illustrated in 
Fig.~9, which shows the coordinate differences from two consecutive 
exposures. It can be reduced only by averaging multiple short
exposures or by extending the exposure time.

\item
CCD mosaic devices require frequent monitoring of their metrics.  For
instance, thermal cycling may trigger a nonelastic change in the
geometry of the CCD chips as it did in the KPNO CCD Mosaic Imager
\citep{pla02}.
\end{enumerate}

\subsection{DAS observations \& reduction techniques}

In previous sections we have shown that a deep high-precision astrometric 
standard is feasible and indeed vital in calibrating the FPA. The NOAO 4m
telescopes and their CCD mosaic imagers are well suited to reach the
desired magnitude range ($10<V<25$) and spatial coverage (10 deg$^{2}$).
The bright end of the magnitude range, $10<V<16$, includes the UCAC stars --
the best existing dense reference frame. In order to cover the DAS
magnitude range and reach stars at $V=25$ with S/N=7 or better, a set
of 10~s, 120~s and 900~s exposures is required at each
telescope pointing.

To achieve the science goals and account for differential color 
refraction (DCR), imaging will be done in Johnson-Cousins $VI$ filters. For
astrometry, it is preferable to observe at longer wavelengths where
atmospheric refraction is smaller, hence $\sim70\%$ of our
observations will be obtained in the $I$ bandpass. 
To reach the {\it accuracy} of 5-15~mas (see Sect. 4.3) across
the entire DAS field and iron out all systematic position-, color-, and
magnitude-dependent errors, multiple passes are required in the $I$
bandpass, each at $0\arcmin$, $6\arcmin$, $12\arcmin$, and 
$18\arcmin$ dithers. This is an
absolute minimum of passes needed to apply the block-adjustment and
self-calibration techniques which are crucial to the success of the
DAS project. Two additional passes are necessary to obtain 
seamless $V$ photometry, which enhances the astrometric precision in
addition to providing star counts. Thus, a total of six complete 
passes per DAS field are required.  In order to obtain absolute
proper motions of all stars relative to faint QSOs and compact
background galaxies, the same sequences should be repeated after 
3-4 years.

The large number of frames -- on the order of 700 per
DAS field at one epoch -- compel usage of unsupervised
reduction techniques. A variant of DOPHOT \citep{sch93} with variable 
point-spread function, developed at NOAO, is our choice in obtaining
the pixel coordinates for all objects in the DAS fields. 
Custom-built software to analyze astrometric CCD data
(L. Winter, private communication) would be used at the US Naval
Observatory to process the same frames.  Thus, similar to Hipparcos,
a two-team effort will ensure quality control and provide the means 
to identify any weaknesses in the reductions. 

Two high-precision local astrometric standards -- NGC 188 in the northern 
hemisphere \citep{pla02} and $\omega$ Cen in the south \citep{lee00,pla05} 
will be used to derive the chip constants and refine the distortion
coefficients.  These parameters and improved look-up tables of 
higher-frequency distortions (mainly due to non-planarity and small tilt 
of the CCD chips and wave-front errors in the optical elements) 
will then fully describe each chip's metric.
Distortion free Cartesian coordinates $x,y$  then can be 
bootstrapped using a direct plane-to-plane transformation \citep{mak04}
which allows us to obtain a `superplate' of the desired size, e.g.,
10 deg$^{2}$. The final step is to convert the superplate's Cartesian
coordinates into the UCAC \citep{zac04} via a robust linear plate model, 
thus minimizing possible local spatial errors in the ICRS representation 
by UCAC. The expected accuracy of the final coordinates should be in the 
range of 5-10~mas, if a star is observed at least six times with 
optimal exposure.
This is 3 to 10 times better than any existing large positional catalog
can offer.  Up to 50\% of an overlap between the frames 
should further strenghten the reliability of superplate coordinates 
by applying the block-adjustment \citep{zac92} and the so-called residual 
technique successfully tested  on the HST WFPC2 camera \citep{and03}.

\subsection{Linking to ICRF}

As described in Sect.~2 and shown on the actual application in Sect.~4.2.2
the uncertainties in linking a catalog to the ICRS are far larger than the
precision level would indicate.  There is a way to substantially reduce 
these uncertainties if we are able to translate the superplate 
directly into the International Celestial Reference Frame
(ICRF). The positional accuracy of individual extragalactic radio sources
defining the ICRF is $\sim0.25$~mas \citep{fey04}. Unfortunately, a very
low sky density of these sources ($<1000$ over the entire sky) currently 
prevents a direct link to the ICRF.  We propose a local densification of
the ICRF in the direction of the DAS fields. That can be done, for example,
by selecting strong sources
from the NRAO VLA Sky Survey (NVSS) at 1.4~GHz \citep{con98}. 
Thus, in the GOT field (see Sect. 3.3), the NVSS contains 26 strong sources 
($S\gtrsim60$~mJy), which are expected to be predominantly 
classical radio galaxies and QSOs. 
One source in this field, J0603+2159, has already made the list of VLBA 
Calibrator Survey \citep{fom03}. More detailed VLBI observations
in the standard S and X bands should provide high-accuracy (1-5~mas)
positions of the NVSS strong sources. The low spatial
resolution of the NVSS survey at $\Theta=45\arcsec$ FWHM does not yield
information on the source structure, hence the real number of useful point
sources is unknown.  The optical counterparts of point sources should
provide the reference fiducials to the ICRF. If the number of these
fiducials is approaching, say, ten, a direct solution into the ICRF is
quite feasible. In the case of a lower number of reference fiducials, we
are limited to zeropoint differences only. In either case the external
accuracy will be constrained much better than just by using the faint
end of the UCAC. 

It should be mentioned that a high-accuracy link to QSOs requires
a high-precision correction for differential color refraction (DCR). The
main source of uncertainty in DCR is a marked difference between 
the spectral energy distributions (SED) for stars and QSOs. To derive
 SED of QSOs,  low-resolution spectrophotometry of their optical 
counterparts is highly desirable. 

\section{Conclusions and recommendations}

The main goal of this paper is to provide a rationale for 
Deep Astrometric Standards. We have shown that DAS are vital for
the next generation of imaging telescopes designed to map out 
the 3-dimensional mass distribution in the Universe. The
underlying very subtle effect of galaxy weak lensing requires
that the  Focal Plane Array be astrometrically extremely well calibrated. 
Establishing the DAS is the only reasonably way at this time to perform 
such calibrations at the level of a 5-10~mas precision.

Although the DAS concept primarily serves the needs of
astrometric calibrations for large telescopes, full consideration 
is given to other benefits that such fields can provide 
to astronomy. The selection of the first DAS fields is well-optimized
to the needs of studies of Galactic structure and kinematics,
especially for the thick and thin Galactic disks. 

Since the minimum amount of  observing time to create a single
DAS field at one epoch with existing CCD mosaic imagers is about 100~hrs, 
the DAS initiative may require an international collaboration among
observatories that have at least a 3m aperture telescope and an imager
with detector collecting area covering $\gtrsim0.25$ deg$^2$. 
If there is enough interest in the community, the same fields can also 
serve as faint photometric standards, but that would require additional 
imaging in the desired bandpasses.  To acquire a proper-motion
component of the DAS, the imaging should commence as soon as possible
and then be repeated 3-4 years later.  The targeted accuracies are 
5-10~mas in positions and 2~mas~yr$^{-1}$ in proper motions down to V=25.
The DAS will be the only faint and accurate standards in the 
pre-GAIA period and may serve well over 1-2 decades.




\acknowledgments

The authors thank David Monet for supplying object coordinates for
the selected archival frames taken with the Subaru Suprime-Cam imager.
One of the authors, I.P., thanks Megan Sosey for the excellent observations 
she obtained for us with the KPNO 4m telescope and Aaron Romanowsky 
for observing the Sgr field at the CTIO 4m telescope. We thank the referee for  
thoughtful comments and questions.  This study is based in part on data 
collected at the Subaru telescope and obtained from the SMOKA science 
archive at the Astronomical Data Analysis Center, which is operated by the 
National Astronomical Observatory of Japan. This work has been supported
in part by National Science Foundation grant AST 04-06689 to 
Johns Hopkins University (I. P.) and by NASA grant HST-AR-09958.01-A
(RFGW and I. P.).
awarded by the Space Telescope Science Institute, which is operated
by the Association of Universities for Research in Astronomy, Inc.,
under NASA contract NAS5-26555.

\clearpage

\begin{figure}
\epsscale{0.8}
\plotone{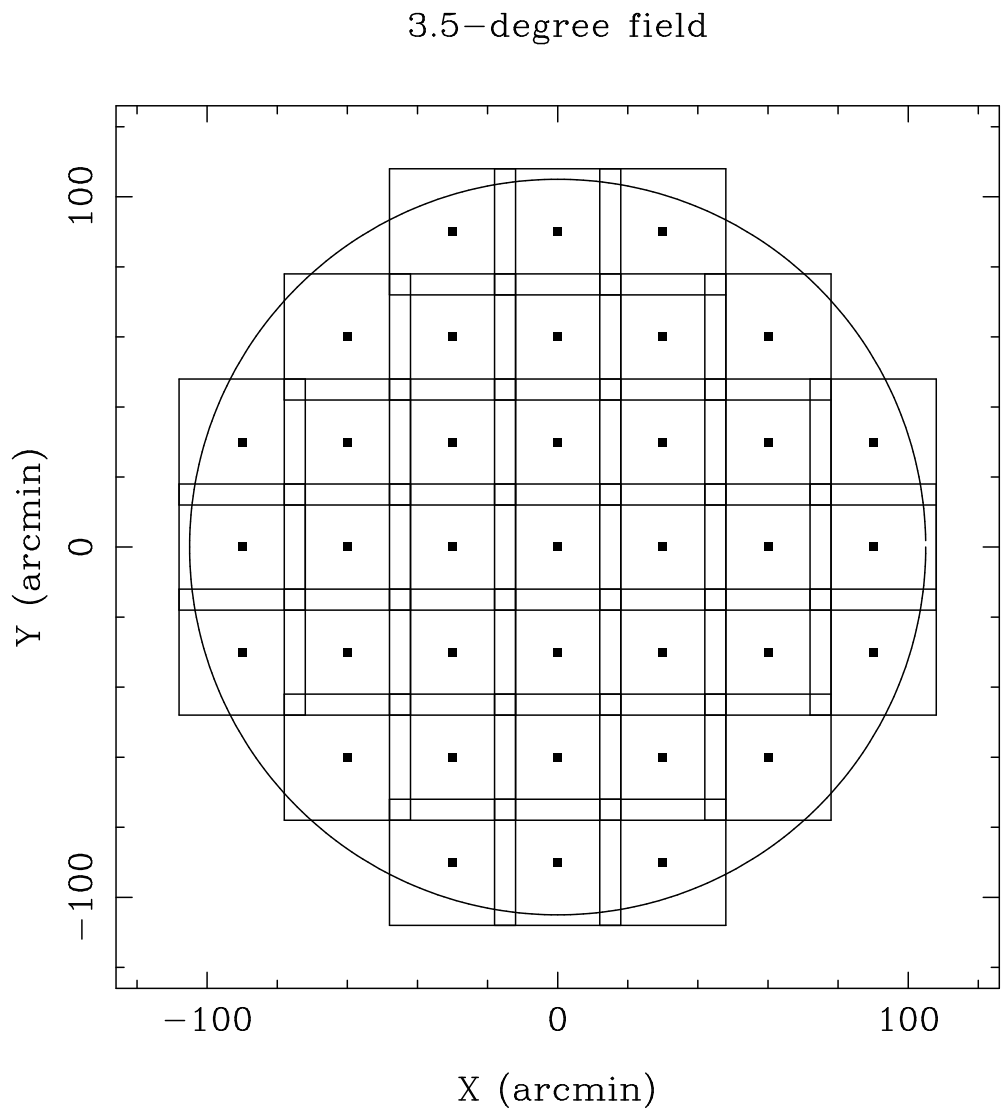}
\caption{Proposed pointing layout in a DAS field with the NOAO 4m telescopes 
and CCD Mosaic imager. In total, 37 partially overlapping ($\Delta=6\arcmin$) 
pointings are necessary to fill in a $3\fdg5$-diameter field. 
The missing parts of the field are filled in by other sets of pointings
at large dithers.}
\end{figure}

\begin{figure}
\epsscale{0.8}
\plotone{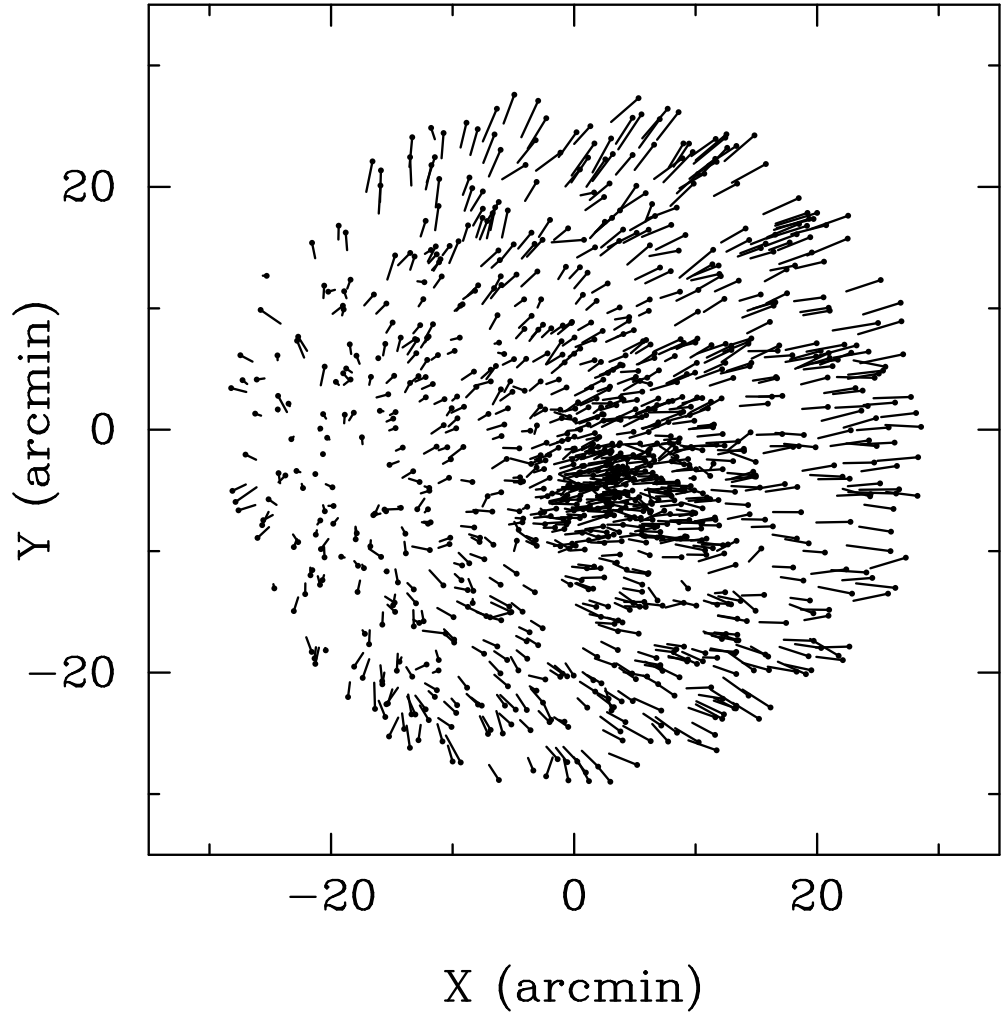}
\caption{Vectorial differences, UCAC--WOCS, around the open cluster
NGC 188. The $XY$ axes represent the gnomonic projection of equatorial
coordinates with a tangent point at RA=$0^{\rm h}$44$^{\rm m}$20$^{\rm s}$
Dec=$+85\arcdeg18\farcm9$ (J200). The largest vector is about 300 mas long.}
\end{figure}

\begin{figure}
\epsscale{0.8}
\plotone{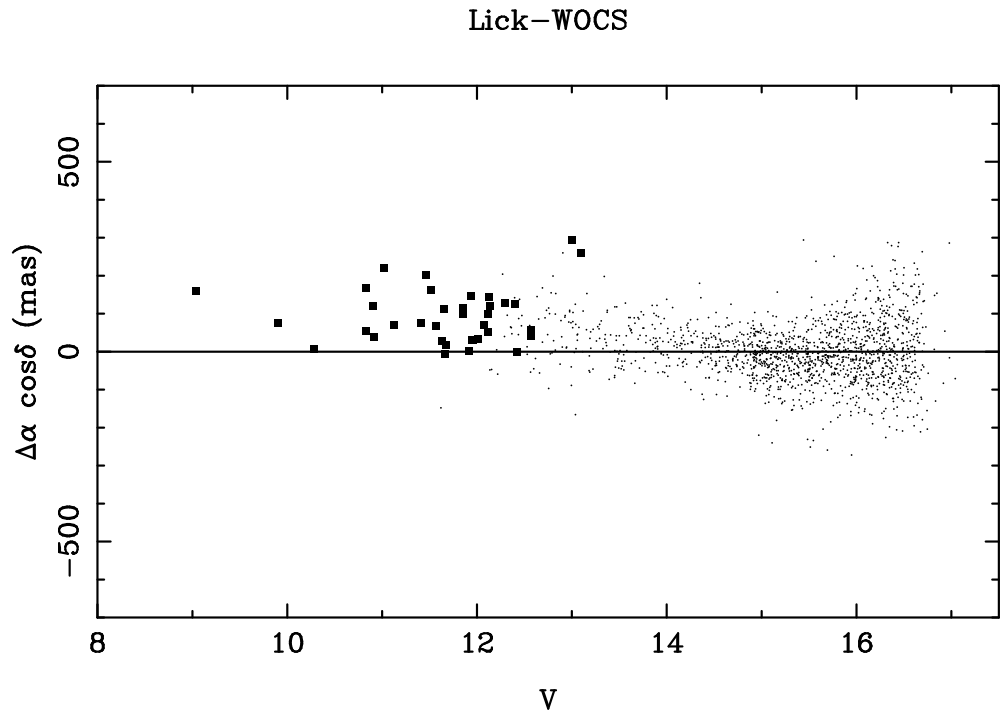}
\caption{Right ascension differences, Lick--WOCS, around 
the open cluster
NGC 188 \citep{pla03}. The filled squares indicate the common Tycho-2
stars. In part, the mean offset of these Tycho-2 stars is a measure of 
the deviation from the ICRS.} 
\end{figure}

\begin{figure}
\epsscale{0.8}
\plotone{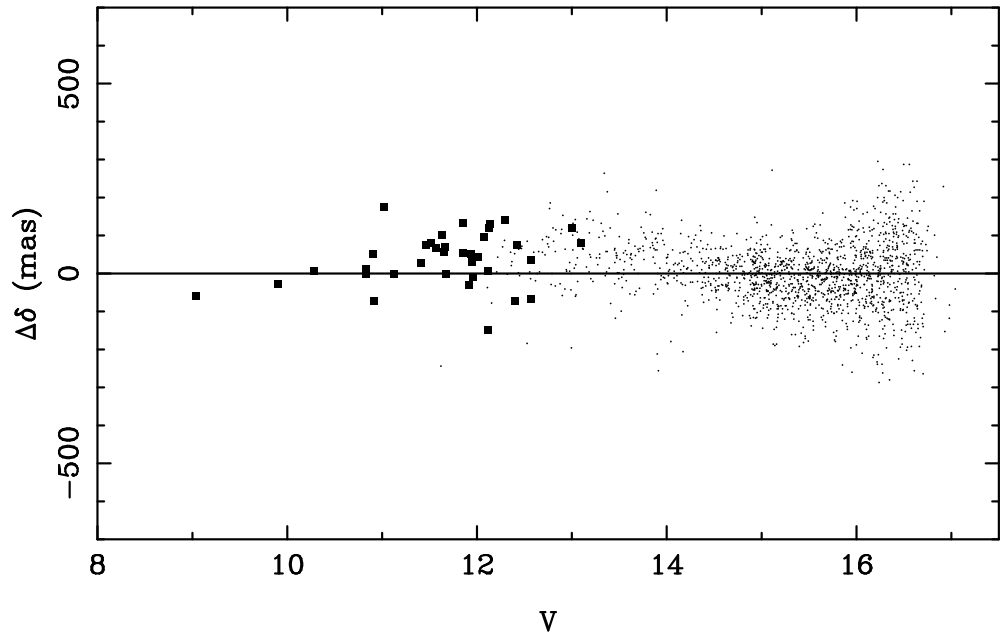}
\caption{Declination differences, Lick--WOCS, around 
the open cluster
NGC 188 \citep{pla03}. The filled squares indicate the common Tycho-2
stars.}
\end{figure}

\begin{figure}
\epsscale{0.8}
\plotone{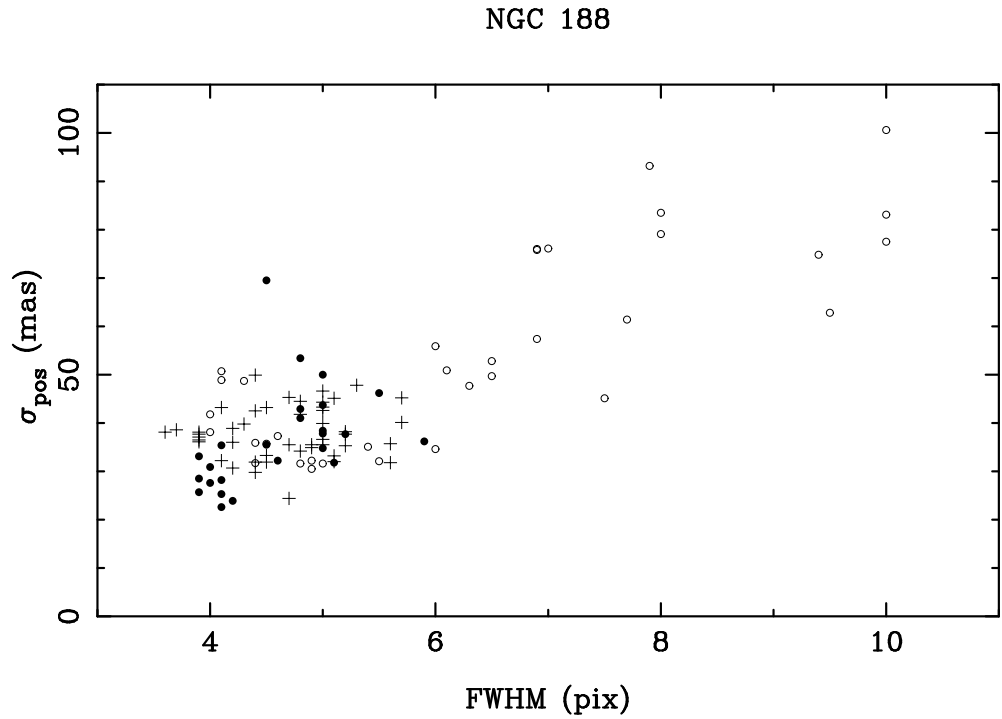}
\caption{Distribution of mean positional dispersions, $\sigma_{\rm pos}$,
as a function of FWHM and exposure time. The data around NGC 188 
are from the NOAO CCD
Mosaic Imager at the Kitt Peak 4m telescope (1 pixel$= 0\farcs26$).
The open circles denote 10-15 s exposures; bold dots -- 30 s;
crosses -- 120-180 s. In this high zenith distance field ($z>53\arcdeg$)
the dependence on exposure time is minimal.}
\end{figure}

\begin{figure}
\epsscale{0.8}
\plotone{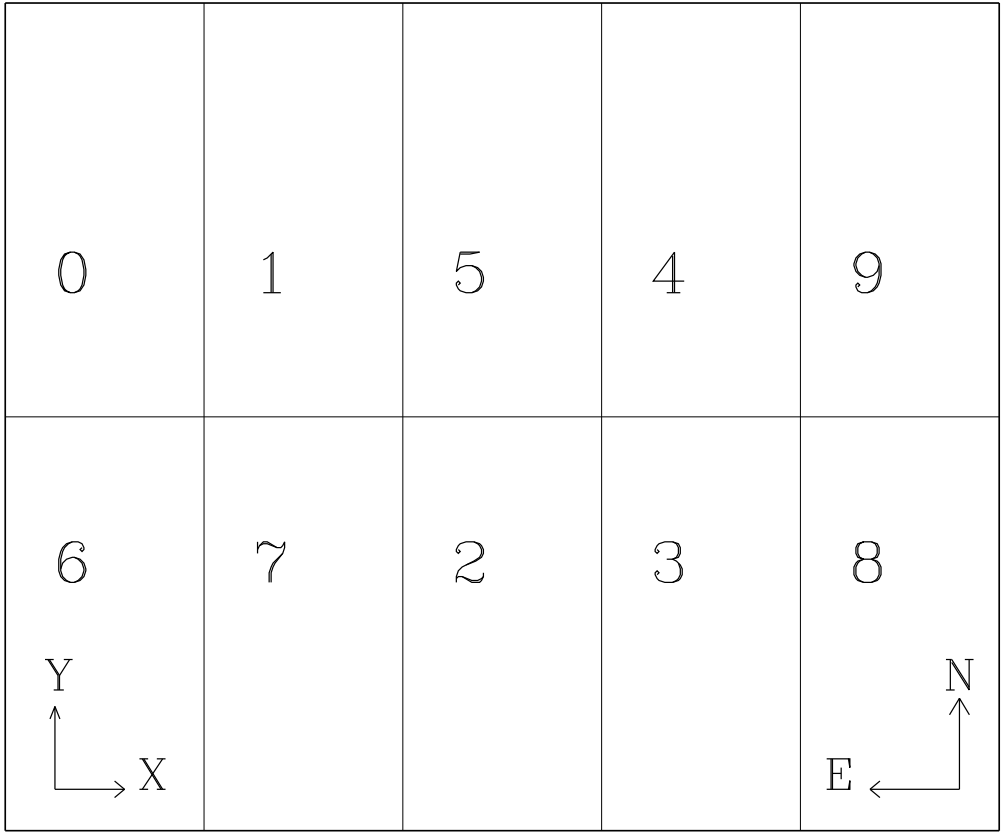}
\caption{Layout of the Subaru Suprime-Cam CCD mosaic imaging plane,
showing the adopted chip numbers, the original pixel coordinate axes 
and the N-E direction.}
\end{figure}

\begin{figure}
\plotfiddle{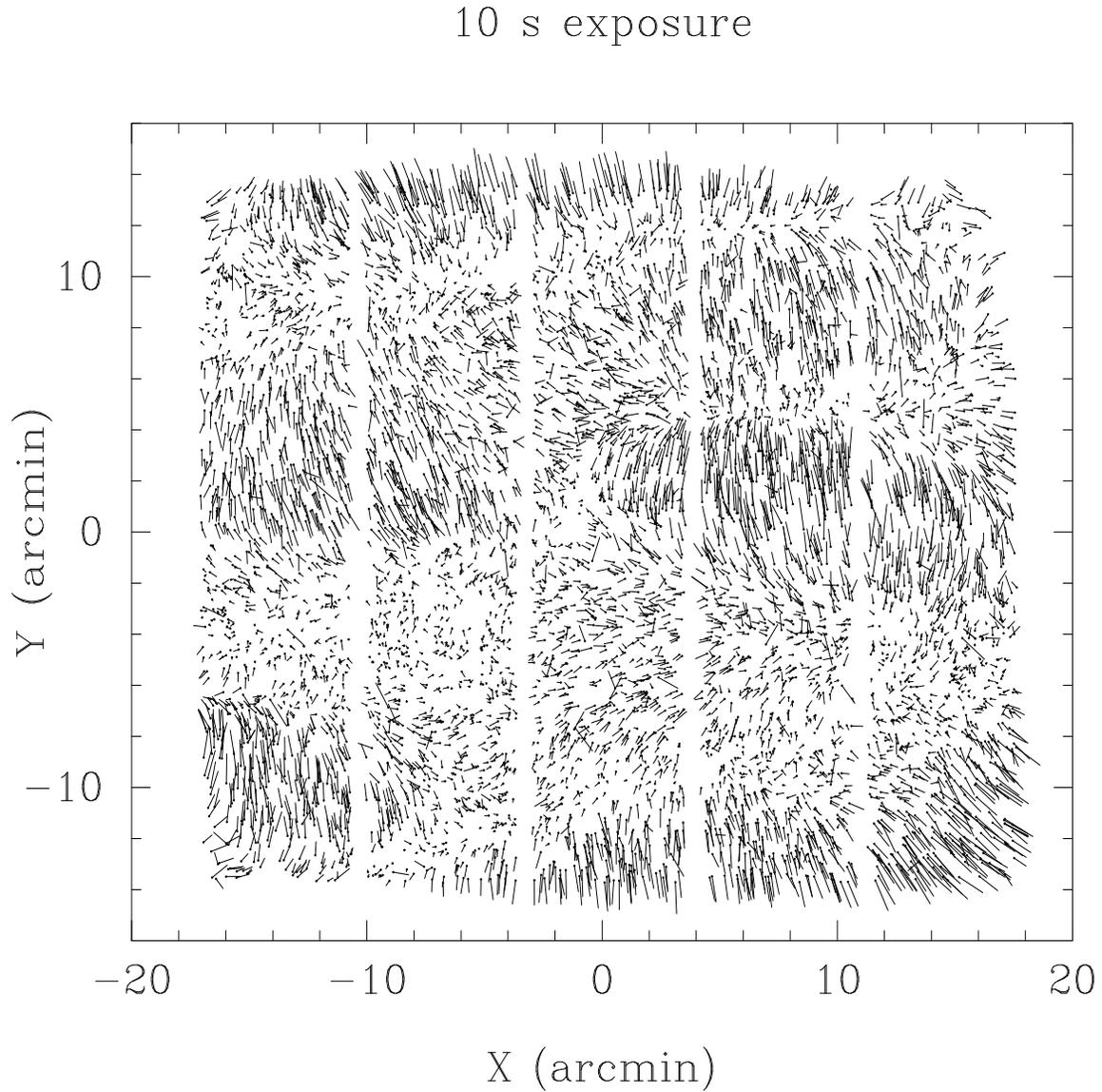}{72pt}{270}{432}{432}{0}{0}
\caption{Vectorial coordinate differences, catalog--CCD frame, for 
the image with the Subaru Suprime-Cam.  
The $XY$ axes represent the gnomonic projection of 
equatorial coordinates.  The selected 10 s exposure shows a correlated
pattern of residuals, mostly due to atmospheric noise. 
The longest vector represents 50 mas.}
\end{figure}

\begin{figure}
\plotfiddle{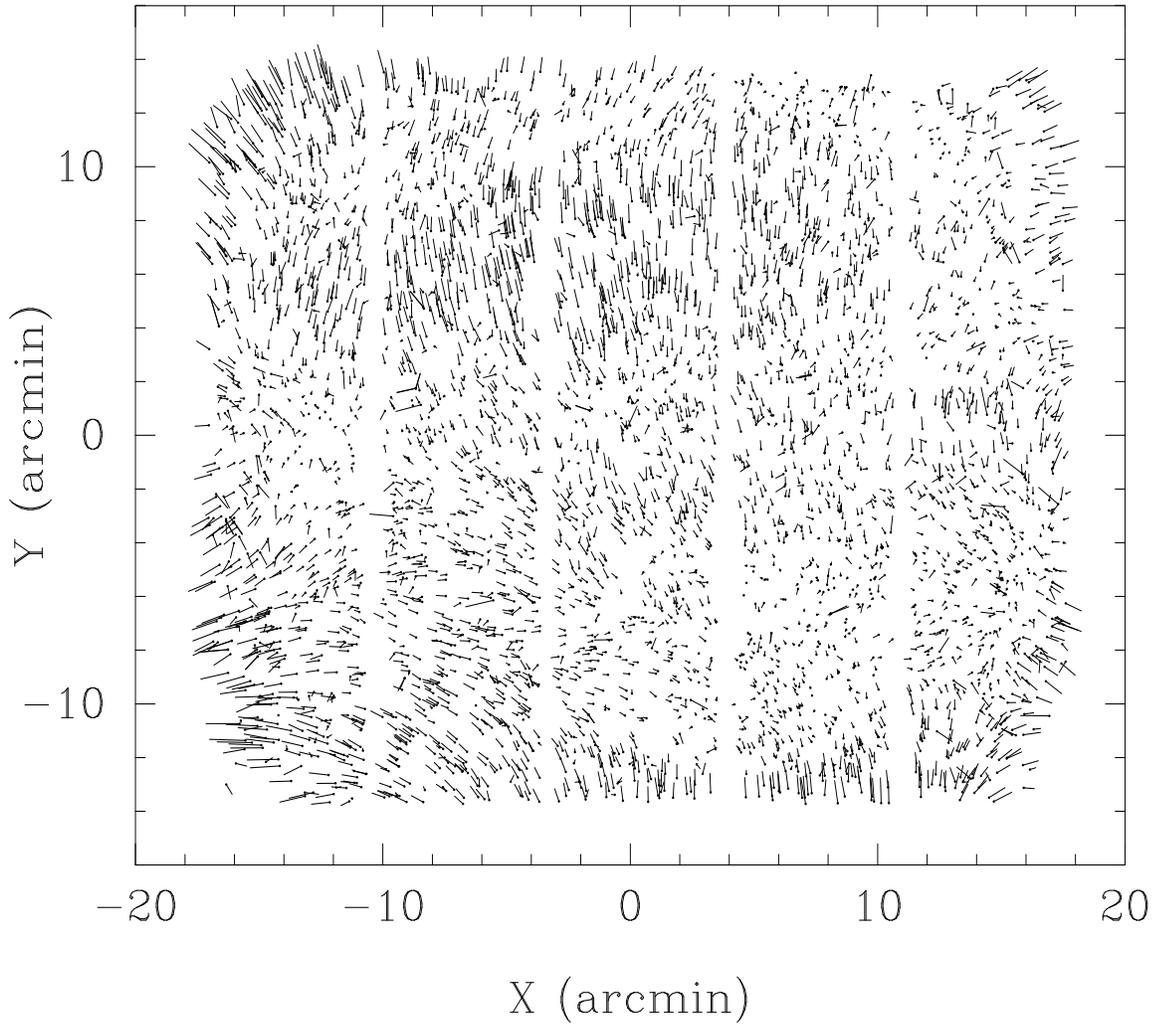}{72pt}{270}{432}{432}{0}{0}
\caption{Vectorial coordinate differences, catalog--CCD frame, for 
the image with the Subaru Suprime-Cam.  
The $XY$ axes represent the gnomonic projection of 
equatorial coordinates.  The distribution of vectors is much smoother 
than in Fig.~7, although some unmodelled field distortions may be seen 
at the field edges. The same scale as in Fig.~7.}
\end{figure}

\begin{figure}
\epsscale{0.8}
\plotone{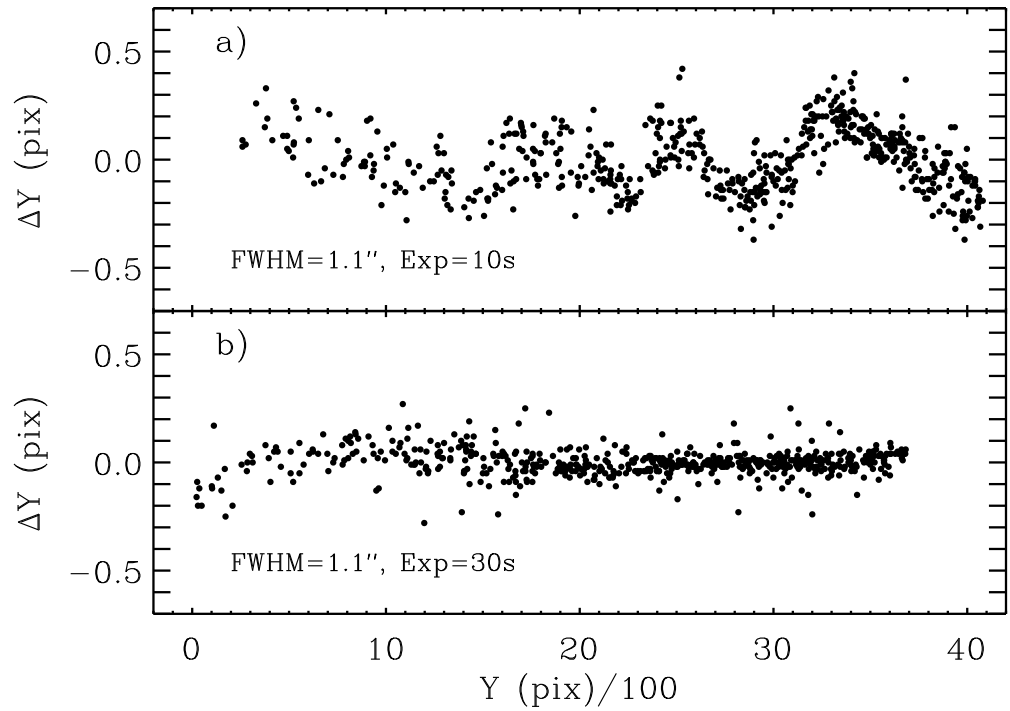}
\caption{Coordinate differences in one NOAO CCD Mosaic Imager chip from two 
consecutive, slightly dithered, exposures. (a) The wiggles in the 10 s
exposures are mostly due to atmospheric noise. (b) In the 30 s exposures
the contribution of atmospheric noise is minimal. (Adapted from
Platais et al. 2002.)}
\end{figure}

\clearpage
\begin{deluxetable}{ccrrrrrrcccc}
\tablewidth{6.5in}
\tablenum{1}
\pagestyle{empty}
\small
\tablecaption{First DAS fields}
\tablehead{
 \colhead{ } & \colhead{Field} & \colhead{RA\tablenotemark{b}} & 
 \colhead{Dec\tablenotemark{b}} &
 \colhead{$\ell$} & \colhead{$b$}   &  
 \colhead{$\lambda$}  &  
 \colhead{$\beta$}& \colhead{$E(B-V)$} &
 \colhead{$n_{\rm U}$\tablenotemark{c}} & 
\colhead{$n_{\rm B}$\tablenotemark{d}} & \colhead{$n_{\rm G}$\tablenotemark{e}}
          }
\startdata
       & GOT\tablenotemark{a} &6:00 & +21:45 & 187.8 & $-0.9$ & 90.0 &$-1.7$ & 
   1.4 & 2260 & 98K & 192K\\ 
       & Hya &8:49 &$ -15$:25 & 241.2 &+17.4 & 139.7 & $-31.9$ &  0.1 & 
  1440 & 31K & 28K  \\ 
       & Oph &17:44 &+11:15 &  35.6 &+34.6 & 265.2 & +34.6 &0.2 & 1510 & 
   29K & 22K \\ 
       & Sgr &19:20 &$-20$:40 &  17.2 &$-15.3$ & 288.7 & +1.5 & 0.1 & 
  3720 & 360K & 215K \\ 
\enddata
\tablenotetext{a}{GOT stands for Gem-Ori-Tau, denoting the constellations
covered in part.}
\tablenotetext{b}{Equatorial coordinates (J2000) of the field center in
hr, min for RA and in deg, arc-min for Dec; followed by
Galactic, $l$ and $b$, and ecliptic coordinates, $\lambda$ and $\beta$, both 
in decimal deg.}
\tablenotetext{c}{Number of UCAC2 stars per deg$^2$.}
\tablenotetext{d}{Besan\c con model: predicted number of stars per deg$^2$ 
down to $V=25$ (in thousands).}
\tablenotetext{e}{Gilmore model: predicted number of stars per deg$^2$ 
down to $V=25$ (in thousands).}
\end{deluxetable}

\clearpage
\begin{deluxetable}{ccrrrccc}
\tablewidth{5.5in}
\tablenum{2}
\pagestyle{empty}
\small
\tablecaption{Suprime-Cam CCD chip constants}
\tablehead{
 \colhead{ } & \colhead{Chip} & \colhead{$dx$} & \colhead{$dy$} &
 \colhead{$\Theta$} & \colhead{$\epsilon_{dx}$}   &  
 \colhead{$\epsilon_{dy}$}  &  \colhead{$\epsilon_{\Theta}$}  
         }
\startdata
       & 0 &-38.53 &  43.32 & -0.004197 & 0.040 & 0.063 & 0.000023\\ 
       & 1 & 39.46 &  38.82 & -0.002558 & 0.022 & 0.046 & 0.000021\\ 
       & 2 &108.28 &   2.48 & -0.002586 & 0.027 & 0.042 & 0.000015\\ 
       & 3 &186.38 &  -5.09 & -0.002277 & 0.017 & 0.027 & 0.000025\\ 
       & 4 &194.42 &  29.84 & -0.004793 & 0.020 & 0.043 & 0.000018\\ 
       & 5 &183.50 &  31.48 & -0.003079 & 0.026 & 0.038 & 0.000010\\ 
       & 6 & 17.64 &   9.41 & -0.002141 & 0.011 & 0.060 & 0.000027\\ 
       & 7 & 96.54 &   4.14 & -0.002591 & 0.025 & 0.036 & 0.000021\\ 
       & 8 &330.08 &  -8.88 & -0.003778 & 0.023 & 0.064 & 0.000020\\ 
       & 9 &271.17 &  24.30 & -0.003386 & 0.043 & 0.064 & 0.000014\\ 
\enddata
\end{deluxetable}

\end{document}